\documentclass[12pt]{article}
\usepackage{amssymb}

\newcommand{\be}{\begin{equation}}
\newcommand{\ee}{\end{equation}}
\def\bea{\begin{eqnarray}}
\def\eea{\end{eqnarray}}

\newcommand{\bn}{\begin{eqnarray}}
\newcommand{\en}{\end{eqnarray}}

\newcommand{\p}{\partial}

\newcommand{\nn}{\nonumber}

\newcommand{\tmm}{\tilde{m}}

\newcommand{\no}{\noindent}

\newcommand{\s}{\,\,\,\,}
\def\bea{\begin{eqnarray}}
\def\eea{\end{eqnarray}}

\newcommand{\beq}{\begin{eqnarray}}
\newcommand{\eeq}{\end{eqnarray}}
\begin{document}

\title{\textbf{Weyl and transverse diffeomorphism invariant spin-2 models in
$D=2+1$ }}
\author{D. Dalmazi$^{1,2}$\thanks{dalmazi@feg.unesp.br}, A.L.R. dos Santos$^{1}$\thanks{alessandroribeiros@yahoo.com.br},
 Subir Ghosh$^{3,2}$\thanks{subirghosh20@gmail.com}, E. L. Mendon\c ca$^{1,2}$\thanks{eliasleite@feg.unesp.br},  \\
\textit{{1- UNESP - Campus de Guaratinguet\'a - DFQ} }\\
\textit{{Guaratinguet\'a, SP Brazil 12516-410.} }\\
\textit{{2- ICTP South American Institute for Fundamental Research, IFT-UNESP} }\\
\textit{{ S\~ao Paulo, SP Brazil  01440-070.} }\\
\textit{{3- Physics and Applied Mathematics Unit, Indian Statistical Institute} }\\
\textit{{ 203 B. T. Road, Kolkata 700108, India} }\\
}
\date{\today}
\maketitle

\begin{abstract}

There are two covariant descriptions of massless spin-2 particles
in $D=3+1$ via a symmetric rank-2 tensor: the linearized
Einstein-Hilbert (LEH) theory and the Weyl plus transverse
differomorphism (WTDIFF) invariant model. From the LEH theory one
can obtain the linearized New Massive Gravity (NMG) in $D=2+1$ via
Kaluza-Klein dimensional reduction followed by a dual master
action.  Here we show that a similar route  takes us from the
 WTDIFF model to a linearized scalar tensor NMG which belongs to a
larger class of consistent spin-0 modifications of NMG. We also
show that a traceless master action applied to a parity singlet
furnishes two new spin-2 selfdual models.

Moreover, we examine the singular replacement $h_{\mu\nu} \to
h_{\mu\nu} - \eta_{\mu\nu}h/D$ and prove that it leads to
consistent massive spin-2 models in $D=2+1$. They include
linearized versions of unimodular topologically massive gravity
(TMG) and unimodular NMG. Although the free part of those
unimodular theories are Weyl invariant,  we do not expect any
improvement in the renormalizability. Both the linearized K-term
(in NMG) and the linearized gravitational Chern-Simons term (in
TMG) are invariant under longitudinal reparametrizations $\delta
h_{\mu\nu} = \p_{\mu}\p_{\nu}\phi$ which is not a symmetry of the
WTDIFF Einstein-Hilbert term. Therefore, we still have one degree
of freedom whose propagator behaves like $1/p^2$ for large
momentum.

\end{abstract}

\newpage

\section{ Introduction}

The covariant description of massless spin-2 particles is very constrained, see for instance \cite{abgv,blas}.
By far the most popular model is the massless limit of the massive Fierz-Pauli (FP) theory \cite{fp}. It is
equivalent to the LEH theory.\footnote{Throughout this work we use $\eta_{\mu\nu}=diag(-,+,\cdots , +)$ and
$\bar{h}_{\mu\nu}\equiv h_{\mu\nu} -
 \eta_{\mu\nu} h /D$. Moreover we use the acronyms LEH, DIFF, TDIFF  and WTDIFF, standing for linearized Einstein-Hilbert,
diffeomorphisms, transverse diffeomorphisms and Weyl and
transverse diffeomorphisms respectively.} It is invariant under
linearized diffeomorphisms $\delta h_{\mu\nu} = \p_{\mu} \xi_{\nu}
+ \p_{\nu} \xi_{\mu}$. The second way is the WTDIFF model, (see
\cite{abgv} and \cite{u,hu} for  earlier references), which is
invariant under linearized diffeomorphisms and Weyl
transformations, i.e., $\delta h_{\mu\nu} = \p_{\mu} \xi_{\nu}^T +
\p_{\nu} \xi_{\mu}^T + \eta_{\mu\nu} \, \phi $ where
$\p^{\mu}\xi_{\mu}^T=0$. The WTDIFF model is the linearized
truncation of unimodular gravity
 \cite{einstein,alvarez1,agmhvm,agmm} which, on its turn, corresponds to the Einstein-Hilbert
 theory with the replacement $ g_{\mu\nu} \to \hat{g}_{\mu\nu}/(- \hat{g})^{1/D} $.

 The WTDIFF model can be obtained from the usual LEH
 theory by the singular replacement $h_{\mu\nu} \to h_{\mu\nu} -
 \eta_{\mu\nu} h /D$. The reason why this replacement is
 successful is not obvious. An argument is given
 in \cite{bfh}. Namely, we first introduce a harmless
 Stueckelberg scalar field altogether with a trivial Weyl
 symmetry in the LEH model via $h_{\mu\nu} \to h_{\mu\nu}
 + \eta_{\mu\nu} \phi $, thus defining a conformal model.
 The Weyl symmetry can be broken by fixing the
 unitary gauge $\phi =0$ which leads us back to LEH. If
 we alternatively choose the gauge $\phi = - h/D$, we keep
 the Weyl symmetry unbroken but the DIFF symmetry is reduced to TDIFF.
 Therefore, the LEH and the WTDIFF models correspond to two different gauges of
 the same conformal theory. This is not a rigorous proof of equivalence since the gauge
 fixing is implemented at action level. According to
 \cite{mst}, the equivalence between a general gauge
 theory and its gauge fixed version (at action level) requires that the
 gauge condition be complete, which is not the case here. The
 key point is that the gauge fixed action leads to less
 equations of motion. This is not equivalent in general to first derive
 the full set of equations of motion and fix the gauge afterwards.

Mainly motivated by the accelerated expansion of the universe, but also as a matter of principle, we are
interested here in massive gravitational theories. They have been a subject of intense work in the last decade,
(see \cite{drgt,hr} and the review works \cite{hinter,drham}). The modern massive gravities are built up on the
top of the massive FP model, so we might wonder whether massive WTDIFF models do exist or even before that, we
must search for WTDIFF massive spin-2 theories. Naive addition of mass terms to the massless WTDIFF model breaks
 unitarity \cite{blas}. In \cite{bfh} the reader can find a recent discussion in that direction via
dimensional reduction.

Notice that the argument of \cite{bfh} can not be used in order to derive a WTDIFF version of the ${\bf
massive}$ FP model. The second gauge $\phi = - h/D$ is not
 allowed, since there is no vector symmetry to be
 fixed or partially fixed in the massive case.
  The best we can do is to replace  $h_{\mu\nu} \to h_{\mu\nu}
 + \eta_{\mu\nu} \phi $ and carry out a field redefinition $\phi
 \to \phi - h/D $. We end up with the  Weyl symmetry $\delta h_{\mu\nu}
 = \eta_{\mu\nu} \Lambda $ but $\phi $ is not pure gauge anymore.
 It remains in the theory as an extra degree of freedom
 \cite{bfh}. This is the typical situation for massive WTDIFF models,
 extra fields are required in general.

 In $D=2+1$ the situation is different. We
 may have massive spin-2 models still invariant under vector
 symmetries. This is the case of the second, third and fourth order selfdual models of helicity +2 or -2 (parity singlets) and
 the linearized version of the new massive gravity (NMG) \cite{bht} with both helicities $\pm$2 (parity doublet).
 This raises the question of
 defining WTDIFF versions of those models according to the argument of
 \cite{bfh} and eventually building up unimodular versions of the
 corresponding massive gravitational theories. This issue is
 specially interesting from the point of view of
 renormalizability because  the highest
 derivative term of topologically massive gravity (TMG) and of NMG
  is Weyl invariant at linearized level, contrary to the lower
  derivative term (Einstein-Hilbert). It would be
  interesting to have both terms Weyl invariant in order to make
  sure that all degrees of freedom have their large momentum
  behavior ruled by the highest derivative term.
  We examine that question here.

 In section II we show the consistency of the WTDIFF version of the linearized NMG
 model and of the second, third and fourth order
 spin-2 selfdual models $SD_n$ as well and comment on possible
 unimodular massive gravities and the issue of renormalizability.
In section III, by means of a traceless master action approach
 we derive a new scalar-tensor selfdual model of second order  $NSD_2$
 and also a new fourth order model $NSD_4$ from $NSD_2$.
 In section IV a traceless master action gives rise to a new
 scalar-tensor NMG model which is shown to be a specific subcase of a
 more general class of consistent spin-0 (scalar tensor) deformations of NMG.
 In section V we present our
 final comments.


%
%
%
%
%
%

\section{WTDIF invariant models in $D=2+1$}

\subsection{$m=0$}

In order to point out the subtleties of gauge fixing procedure at action level, it is instructive to first
look at the massless case. It is known that the Einstein-Hilbert theory has no propagating degrees of freedom in
$D=2+1$. At linearized level we have:

\be S_{LEH} [h_{\mu\nu}] = \int \, d^3 x \, \left( \sqrt{-g} R
\right)_{hh} = (1/4) \int \, d^3 x \, h_{\rho\gamma}
E^{\rho\delta}E^{\gamma\sigma} h_{\delta\sigma}  \quad ,
\label{leh} \ee

\no where the transverse operators

\be E^{\rho\delta} \equiv \epsilon^{\rho\delta\sigma}\p_{\sigma} \quad ; \quad  \Box \theta_{\rho\sigma} \equiv
\Box \eta_{\rho\sigma} - \p_{\rho} \p_{\sigma} \label{etheta} \ee

\no are such that

\be  E^{\mu\nu} E ^{\alpha\beta} = \Box \left( \theta^{\mu\beta}\theta^{\nu\alpha} -
\theta^{\mu\alpha}\theta^{\nu\beta} \right) \quad . \label{id1} \ee

\no The equations of motion $ E^{\rho\delta}E^{\gamma\sigma}
h_{\delta\sigma} =0 $ are equivalent (multiply by
$\epsilon_{\rho\mu\nu}\epsilon_{\gamma\alpha\beta}$) to a
vanishing linearized Rieman curvature
$R_{\mu\nu\alpha\beta}^{L}(h)=0$ (flat space). The general
solution is pure gauge $h_{\mu\nu}=\p_{\mu}\xi_{\nu} +
\p_{\nu}\xi_{\mu} $.

On the other hand, if we make the Stueckelberg replacement
$h_{\mu\nu} \to h_{\mu\nu}
 + \eta_{\mu\nu} \phi $ in (\ref{leh}) followed by  the gauge fixing
 $\phi = -h/3$  at the action level, we have a WTDIFF invariant model $
S_{LEH} [\bar{h}_{\mu\nu}]$ whose equations of motion are traceless:

\be E^{\rho\delta}E^{\gamma\sigma} \bar{h}_{\delta\sigma} - \eta^{\rho\gamma} \p^{\mu}\p^{\nu}\bar{h}_{\mu\nu}/3
=0  \quad . \label{weom} \ee

\no Applying $\p_{\rho}$ we show that the linearized scalar curvature is an arbitrary constant, not necessarily
vanishing anymore, i.e., $R^L = \p^{\mu}\p^{\nu}\bar{h}_{\mu\nu}=c$. The integration constant $c$ can not be
fixed by the equations of motion. Contracting (\ref{weom}) with
$\epsilon_{\rho\mu\nu}\epsilon_{\gamma\alpha\beta}$ we have a maximally symmetric space in general, not
necessarily flat:

\be R_{\mu\nu\alpha\beta}^{L}(\bar{h})= \frac c6
(\eta_{\mu\beta}\eta_{\nu\alpha}-\eta_{\mu\alpha}\eta_{\nu\beta})
\quad . \label{mss} \ee

\no   The solution to (\ref{mss}) is given by

\be \bar{h}_{\mu\nu} = \p_{\mu}\xi_{\nu}^T + \p_{\nu}\xi_{\mu}^T
+ \frac c{10} \left( x_{\mu} x_{\nu} - \eta_{\mu\nu} \frac{x^2}3
\right) \quad . \label{gs} \ee

\no Except for the $c$-dependent term, the solution is pure gauge. So the number of propagating degrees of
freedom still vanishes. We have only one (not one  infinity) extra degree of freedom represented by $c$ but the
geometry has been changed as if we had a cosmological constant\footnote{In $D=2+1$ the Rieman tensor is
proportional to the Ricci tensor, see e.g. \cite{weinberg}}. On the other hand, if after the Stueckelberg
substitution in (\ref{leh}) we first derive the field equations with respect to $h_{\mu\nu}$ and $\phi$ and only
afterwards we fix the gauge $\phi = -h/3$ we would have obtained $R_{\mu\nu\alpha\beta}^{L}(\bar{h})=0$ and
consequently $R^L = \p^{\mu}\p^{\nu}\bar{h}_{\mu\nu}=0$ which corresponds to $c=0$. We learn that the gauge
fixing at action level is nontrivial, specially regarding gravitational theories. Thus, whenever we fix a gauge
at action level, as in the next section, we must explicitly check the consistency of the resulting model. We can
not take physical equivalence for granted. In the following subsections we turn to massive models which are
still gauge invariant in $D=2+1$ dimensions.

\subsection{WTDIFF linearized New Massive Gravity (parity doublet)}

The linearized version of the New Massive Gravity \cite{bht} can
be written in the compact Fierz-Pauli form

\bea S_{LNMG}[h] &=& \int \, d^3 x \,  \sqrt{-g} \left\lbrack \frac 1{m^2} \left( R_{\mu\nu}^2 - \frac 38 R^2
\right)- R \right\rbrack_{hh} \quad , \label{nmg} \\
&=& (1/4) \int \, d^3 x \, \left\lbrack h_{\rho\gamma} E^{\rho\delta}E^{\gamma\sigma} h^*_{\delta\sigma} - m^2 (
h^{\mu\nu} h_{\mu\nu}^* - h \, h^*) \right\rbrack \quad , \label{lnmg} \eea

\no where the dual field is given by \cite{sd4}

\be h_{\mu\nu}^*[h] = \frac 1{m^2} \left( E_{\mu\rho}E_{\nu\sigma}h^{\rho\sigma} + \frac 12 \eta_{\mu\nu}\Box
\theta_{\rho\sigma}h^{\rho\sigma} \right) \quad , \label{hdual} \ee

\no and identically satisfies

\be \p^{\mu}h_{\mu\nu}^* = \p_{\nu}h^*  \quad . \label{id2} \ee

\no  If we replace $h_{\mu\nu}^*$ by $h_{\mu\nu}$ in (\ref{lnmg})
we recover the usual massive FP model. The theory $S_{LNMG}[h]$ is
DIFF invariant. Repeating in (\ref{lnmg}) the procedure of the
last subsection, which amounts to the replacement $h_{\mu\nu} \to
\bar{h}_{\mu\nu}$ at action level, we derive a WTDIFF version of
the linearized NMG: $S_{WLNMG} (h) = S_{LNMG} (\bar{h})$. Let us
check the particle content of $S_{WLNMG}$. The equations of motion
$\delta S_{WLNMG}/\delta h^{\mu\nu} =0$ are traceless, namely,

\be
E_{\mu}^{\,\,\rho}E_{\nu}^{\,\,\sigma}h^*_{\rho\sigma}[\bar{h}]+
\frac 13 \eta_{\mu\nu} \Box \theta^{\rho\sigma}h^*_{\rho\sigma} =
m^2 \, \left(h^*_{\mu\nu}[\bar{h}] - \frac 13\eta_{\mu\nu}
h^*[\bar{h}]\right) = m^2\, \bar{h}^*_{\mu\nu}[\bar{h}]  \quad .
\label{wnmgeom} \ee

\no From (\ref{id2}) we see that $\Box
\theta^{\rho\sigma}h^*_{\rho\sigma}=0$. Applying $\p^{\mu}$ on
(\ref{wnmgeom}) we have $\p^{\mu}\bar{h}^*_{\mu\nu}[\bar{h}]=0$.
Using the identities (\ref{id1}) and (\ref{id2}) we see that
(\ref{wnmgeom}) is equivalent to the Klein-Gordon equations $(\Box
- m^2)\bar{h}_{\mu\nu}^*[\bar{h}] =0$. Therefore,
$\bar{h}^*_{\mu\nu}[\bar{h}]$ is transverse, traceless and
satisfies the Klein-Gordon equations. Moreover it is invariant
under the WTDIFF gauge symmetry of $S_{WLNMG}$, i.e., $\delta
h_{\mu\nu} = \p_{\mu} \xi_{\nu}^T + \p_{\nu} \xi_{\mu}^T +
\eta_{\mu\nu} \, \phi $. So $S_{WLNMG}$ correctly describes
massive spin-2 particles. From (\ref{id2}) and
$\p^{\mu}\bar{h}_{\mu\nu}^* =0$ we have $\p_{\mu} h^* =0$, so
$h^*=\eta^{\alpha\beta}h_{\alpha\beta}^*$ becomes an integration
constant which plays no role from the point view of the particle
content of $S_{WLNMG}$ but from the point of view of a linearized
gravitational theory works like a cosmological constant.

Although LNMG can be obtained from the usual massive FP model via
master action, see for instance \cite{sd4}, we have not been able
to derive the WLNMG model from any second order theory via master
action. The WLNMG model contains both helicities $+2$ and $-2$, in
the next subsection we look at parity singlets of helicity $+2$ or
$-2$ described in terms of a symmetric traceless tensor.

%

\subsection{WTDIFF self-dual models (parity singlets)}

Free helicity $+2$ or $-2$ states can be described by the so called spin-2 self-dual models (SDn), of n-th order
in derivatives with $n=1,2,3,4$. The SDn model can be obtained from the SD(n-1) via a consecutive Noether gauge
embedding procedure as shown in \cite{sd4}. The equivalence among all those models can be proved by means of a
master action approach \cite{dj}, see \cite{prd2009,sd4}, which also furnishes a dual map $e_{\mu\nu} \to
e_{\mu\nu}^*$ responsible for the equivalence of correlation functions of $e_{\mu\nu}$ in the SD1 model with
correlation functions of $e_{\mu\nu}^*$ in the higher order SDn models. All SDn models can be written in a
compact way{\footnote{A similar formula holds in the spin-1 case where $n=1,2$. The  SD2 model is the
Maxwell-Chern-Simons theory of \cite{djt} and SD1 was suggested in \cite{tpn}. Namely, \be {\cal
L}_{SDn}^{(1)}=\frac{m}{2}A_{\mu}E^{\mu\nu}A_{\nu}^* - \frac{m^2}2 A^{\mu}A_{\mu}^*\ee \no where
$A_{\mu}^*=A_{\mu}$ for the SD1 case while $A_{\mu}^*=E_{\mu\nu}A^{\nu}/m$ in the SD2 case.}  similar to the
first-order model of Aragone and Khoudeir \cite{aragone}, which was the first one to be suggested, namely

\be {\cal
L}_{SDn}^{(2)}=\frac{m}{2}e_{\mu}^{\s\nu}E^{\mu\alpha}e^*_{\alpha\nu}-\frac{m^2}{2}(e^{\mu\nu}e_{\nu\mu}^*-e\,
e^*)\label{sdn}\ee \no where

\bea e_{\mu\nu}^* (n=1)&=&\, e_{\mu\nu} \nn \\
e_{\mu\nu}^*(n=2)&=&\,\frac{E_{\nu}^{\s\alpha}e_{\alpha\mu}}{m}\,
+ \,\frac{\eta_{\mu\nu}E^{\alpha\beta}e_{\alpha\beta}}{2m}\nn\\
e_{\mu\nu}^* (n=3)&=&\frac{E_{\mu}^{\s\alpha}E_{\nu}^{\s\beta}e_{(\alpha\beta)}}{m^2}
+\frac{\eta_{\mu\nu}\Box\theta^{\alpha\beta}e_{\alpha\beta}}{2m^{2}}\nn\\
e_{\mu\nu}^*
(n=4)&=&\,\frac{(E_{\mu}^{\s\alpha}\Box\theta_{\nu}^{\s\beta}+E_{\nu}^{\s\alpha}\Box\theta_{\mu}^{\s\beta})
e_{(\alpha\beta)}}{2m^3}\eea

\no The equations of motion from (\ref{sdn}) are then given by: \be E_{\mu}^{\s\alpha}e^*_{\alpha\nu}-
m(e^*_{\nu\mu}-\eta_{\mu\nu}e^*)=0\label{PL}\ee

\no  Applying $\partial_{\mu}$ in (\ref{PL}) we have

\be \p^{\nu} e_{\mu\nu}^*  = \p_{\mu} e^* \quad . \label{id3} \ee

\no Notice that (\ref{id3}) holds identically for the higher order cases $n=2,3,4$ as a consequence of a local
vector symmetry in those cases. Next, by acting with $\epsilon_{\mu\nu\lambda}$ on (\ref{PL}) we conclude that
$e^*_{[\mu\nu]}=0$. If we take the trace of (\ref{PL}) we obtain $e=0$. Therefore
$\partial^{\mu}e^*_{\mu\nu}=0$. Then \be E_{\mu}^{\s\alpha}e^*_{\alpha\nu}+E_{\nu}^{\s\alpha}e^*_{\alpha\mu}+
2\, m\, e^*_{(\mu\nu)}=0\label{PL2}\ee Now if we apply $E_{\mu\sigma}$ in (\ref{PL2}) we obtain the Klein-Gordon
equation for $e^*_{(\mu\nu)}$: \be (\Box-m^{2})e^*_{(\mu\nu)}=0\ee \no Therefore ${\cal L}_{SDn}^{(2)}$
represent a massive particle of helicity $+2$ for all cases $n=1,2,3$ and $4$.

The $SDn$ models, with $n=2,3,4$, are invariant under the
following respective gauge transformations: \be \delta_2
e_{\mu\nu}=\p_{\mu}V_{\nu}\quad;\quad \delta_3
e_{\mu\nu}=\p_{\mu}V_{\nu}+\Lambda_{[\mu\nu]}\quad;\quad \delta_4
e_{\mu\nu}=\p_{\mu}V_{\nu}+\Lambda_{[\mu\nu]}+\eta_{\mu\nu}\phi\ee
\no where $\Lambda_{[\mu\nu]}=-\Lambda_{[\nu\mu]}$ stand for
arbitrary antisymmetric shifts. If we replace $e_{\mu\nu}$ by its
traceless part $\bar{e}_{\mu\nu}=e_{\mu\nu}-\eta_{\mu\nu}e/3$ in
${\cal L}_{SDn}^{(2)}$, the models will be invariant under
transverse diffeomorphisms and Weyl transformations i.e.: \be
\delta_{W}e_{\mu\nu}=\p_{\mu}V_{\nu}^T+\eta_{\mu\nu}\phi\ee \no
with $\p^{\mu}V_{\mu}^T=0$. So we can define the models:

\be {\cal L}_{WSDn}^{(2)}(e_{\mu\nu})={\cal L}_{SDn}^{(2)}(\bar{e}_{\mu\nu}) \quad , n=2,3,4\ee

\no which lead us to the following traceless equations of motion: \be
E_{\mu}^{\s\alpha}e^*_{\alpha\nu}(\bar{e})+\frac{\eta_{\mu\nu}}{3}E^{\alpha\beta}e^*_{\alpha\beta}(\bar{e})+
m\Big[e^*_{\nu\mu}(\bar{e})-\frac{\eta_{\mu\nu}}{3}e^*(\bar{e})\Big]=0.\label{eomf}\ee

\no Due to (\ref{id3}) which holds identically for $n=2,3,4$, after applying $\epsilon_{\mu\nu\beta}$ on
(\ref{eomf}) it follows that $e^*_{\alpha\beta}(\bar{h})= e^*_{\beta\alpha}(\bar{h})$ which implies
$E^{\alpha\beta}e^*_{\alpha\beta}(\bar{h})=0$. By applying $\p^{\mu}$ on (\ref{eomf}) we have $\p^{\mu}e^*=0$,
so $e^*$ must be constant. Thus, the trace of the original equations of motion of the usual models ${\cal
L}_{SDn}$, i.e. $e^*=0$ is recovered up to an integration constant. This is typical for WTDiff modifications of
diffeomorphisms invariant theories. Notice however, that (\ref{eomf}) is equivalent to (\ref{PL})   when
$e^*_{\mu\nu}$ is replaced by $\bar{e}_{\mu\nu}^*=e^*_{\mu\nu}-\eta_{\mu\nu}e^*/3$. Consequently, we deduce the
Klein-Gordon equations, the helicity equation (\ref{PL2}) and the Fierz-Pauli conditions which assures that
${\cal L}_{WSDn}$ have the same particle content of the ${\cal L}_{SDn}$ models.

\subsection{A note on renormalizability}

The models $S_{LNMG}$, SD3 and SD4  have gravitational nonlinear
completions, they correspond respectively to NMG, topologically
massive gravity (TMG) and higher derivative topologically massive
gravity (HDTMG). In the cases of $S_{LNMG}$ and SD3 the highest
derivative term, of fourth and third order respectively, is
invariant under WDIFF $\delta h_{\mu\nu} = \p_{\mu}\xi_{\nu} +
\p_{\nu}\xi_{\mu} + \delta_{\mu\nu}\Lambda $ while the lowest
derivative term  (linearized Einstein-Hilbert) is only invariant
under DIFF. As argued in \cite{deserprl} this is an obstruction to
the renormalizability of their nonlinear completions, since there
will always be one metric degree o freedom (absent in the highest
derivative term due to the Weyl symmetry) whose propagator is
governed by the Einstein-Hilbert term and behaves unfortunately
like $1/p^2$ for large momentum.

On the other hand, in the last subsections we have shown that
WLNMG  and WSD3   correctly describe free massive spin-2
particles. They are obtained from LNMG and SD3 by the replacement
$h_{\mu\nu} \to \bar{h}_{\mu\nu} \equiv h_{\mu\nu} -
\eta_{\mu\nu}h/3$ which assures that the Weyl symmetry is present
in all sectors of the Lagrangian. In fact, they are invariant
under WTDIFF transformations. The nonlinear version of such
replacement, i.e., $g_{\mu\nu} \to \hat{g}_{\mu\nu} \equiv
g_{\mu\nu}/(-g)^{1/3}$ leads to unimodular theories $\hat{g}=-1$
which are invariant under Weyl transformations and volume
preserving diffeomorphisms ($\nabla^{\mu}\xi_{\mu}$=0). Now we can
be sure that both highest and lowest derivative terms in the
quadratic part of the action are invariant under WTDIFF by
construction. So we may hope that all degrees of freedom behave
like $1/P^4$ in the case of unimodular NMG or $1/P^3$ for
unimodular topologically massive gravity respectively. However,
there is a subtlety. Due to their Weyl symmetry, the highest
derivative terms are unchanged by the replacement $h_{\mu\nu} \to
\bar{h}_{\mu\nu}$. So, they remain invariant under full WDIFF
while the Einstein-Hilbert term is only invariant under WTDIFF.
Consequently, the linearized K-term (NMG case) and the linearized
gravitational Chern-Simons term (TMG case) still have one more
local symmetry than the EH term, namely, they are invariant under
longitudinal diffeomorphisms: $\delta \, h_{\mu\nu} =
\p_{\mu}\p_{\nu} \zeta$. Indeed, such symmetry can be used in
order to obtain the WSD4 model, which is equivalent to SD4,  from
the WSD3 model via Noether gauge embedding just like the Weyl
symmetry is used to get from SD3 to SD4 as shown in \cite{sd4}.
Therefore, the pure longitudinal sector of the metric will behave
like $1/P^2$. So there is no improvement in the renormalizability
as we go to the unimodular theories.

The case of HDTMG \cite{sd4,andringa}, i.e., the nonlinear
completion of SD4,  is even worse from the point of view of
perturbative quantum field theory. Both terms of the quadratic
(free) piece of HDTMG, i.e., the linearized K-term and the
linearized gravitational Chern-Simons term are invariant under
linearized WDIFF while the cubic and higher vertices are only
invariant under DIFF. Thus, there is one metric degree of freedom
which only appears in the vertices without any free propagator. At
quantum level it gives rise to a nonlinear constraint whose role
is unclear. A similar problem also appears in the massless limit
of NMG as discussed in \cite{deg}. The replacement $g_{\mu\nu}\to
\hat{g}_{\mu\nu}$, which amounts to $h_{\mu\nu} \to
\bar{h}_{\mu\nu} $ in the quadratic (${\cal O}(h^2)$) piece of the
theory, turns the DIFF symmetry into Weyl plus volume preserving
diffeomorphisms or WTDIFF at linearized level. However, the
quadratic theory is invariant under the larger WDIFF
transformations, so the pure longitudinal degree of freedom
($\p_{\mu}\p_{\nu} \zeta $) of the metric only appears in the
vertices leading us to an awkward constraint again. The only hope
is to start with the SD4 model and  examine the addition of cubic
and higher vertices invariant under the full set of WDIFF.

\section{New massive spin-2 models via a
traceless master action}

\subsection{Selfdual models}

Let us consider the first order self-dual model originally proposed by \cite{aragone}:

\be S_{SD1}[f]=\int\,d^3x\,\left\lbrack-\frac{m}{2}f_{\mu\nu}E^{\mu\alpha}f_{\alpha}^{\s\nu}
-\frac{m^2}{2}(f_{\mu\nu}f^{\nu\mu}-f^2)\right\rbrack\ee

\no We can split the non-symmetrical field $f_{\mu\nu}$ into its traceless and trace full parts by making
$f_{\mu\nu}=\bar{f}_{\mu\nu}+\eta_{\mu\nu}\phi$ where $\phi$ is a fundamental scalar field. After that we have:

\be
S_{SD1}[\bar{f},\phi]=\int\,d^3x\,\left\lbrack-\frac{m}{2}\bar{f}_{\mu\nu}E^{\mu\alpha}\bar{f}_{\alpha}^{\s\nu}-m\bar{f}_{\mu\nu}E^{\mu\nu}\phi
-\frac{m^2}{2}\bar{f}_{\mu\nu}\bar{f}^{\nu\mu}+3m^2\phi^2 \right\rbrack\ee

The traceless Chern-Simons like term is invariant under $\delta
\bar{f}_{\mu\nu}=\p_{\mu}\xi_{\nu}^T$ with
$\p^{\nu}\xi_{\nu}^T=0$. Moreover, it is possible to show that it
is trivial, it has no particle content by itself. This fact tells
us that it might be used as a mixing term in order to construct a
master action: \be
S_{M}[\bar{f},\bar{e},\phi]=S_{SD1}[\bar{f},\phi]+\frac{m}{2}\int\,
d^3x\,
(\bar{f}_{\mu\nu}-\bar{e}_{\mu\nu})E^{\mu\alpha}(\bar{f}_{\alpha}^{\s\nu}-\bar{e}_{\alpha}^{\s\nu})\ee

\no Then it might be possible to interpolate between the first-order self-dual model and alternative traceless
descriptions. In order to implement it we define a generating functional by adding a source term to the field
$f_{\mu\nu}$,

\be W[f,\phi]=\int\,{\cal D}\bar{f}_{\mu\nu}{\cal D}\bar{e}_{\mu\nu}{\cal D}\phi\,\exp i\left\lbrace
S_M[\bar{f},\bar{e},\phi]+\int\,d^3x\,\left\lbrack\bar{f}_{\mu\nu}\bar{T}^{\nu\mu}+\phi
T\right\rbrack\right\rbrace\ee

\no where one can easily see that after the shift $\bar{e}_{\mu\nu}\to\bar{e}_{\mu\nu}+\bar{f}_{\mu\nu}$ we have
basically the first order self-dual model, since we end up with a completely decoupled Chern-Simons trivial
term. On the other hand without any shifts, we would have:

\bea
S_{M}[\bar{f},\bar{e},\phi]&=&\int\,d^3x\,\Big[\frac{m}{2}\bar{e}_{\mu\nu}E^{\mu\alpha}\bar{e}_{\alpha}^{\s\nu}-m\bar{f}_{\mu\nu}
E^{\mu}_{\s\alpha}(\bar{e}^{\alpha\nu}+\eta^{\alpha\nu}\phi)\nn\\
&-&\frac{m^2}{2}\bar{f}_{\mu\nu}\bar{f}^{\nu\mu}+3m^2\phi^2 +\bar{f}_{\mu\nu}\bar{T}^{\nu\mu}+\phi{T}\Big]\eea

\no After functionally integrating over $\bar{f}_{\mu\nu}$ and shifting: \be \bar{f}_{\mu\nu}\to
\bar{f}_{\mu\nu}
-\frac{1}{m}E_{\nu}^{\s\lambda}(\bar{e}_{\lambda\mu}+\eta_{\lambda\mu}\phi)-\frac{1}{3m}\eta_{\mu\nu}E^{\lambda\sigma}\bar{e}_{\lambda\sigma}+\frac{\bar{T}_{\mu\nu}}{m^2}\ee

\no we can obtain the alternative second order self-dual model given by:
\bea S_{ASD2}[\bar{e},\phi]&=&\int\,d^3x\,\Big[\frac{m}{2}\bar{e}_{\mu\nu}E^{\mu\alpha}\bar{e}_{\alpha}^{\s\nu}+\frac{1}{2}\bar{e}_{\mu\nu}
\left(E^{\mu\beta}E^{\nu\alpha}+\frac{1}{3}E^{\mu\nu}E^{\alpha\beta}\right)\bar{e}_{\alpha\beta}-\bar{e}_{\mu\nu}\Box\theta^{\mu\nu}\phi\nn\\
&-&\phi(\Box-3m^2)\phi+\bar{e}_{\mu\nu}^*(\bar{e},\phi)\bar{T}^{\nu\mu}+\phi T+{\cal{O}}(\bar{T}^2)\Big]\label{asd2}\eea

\no where we have neglected quadratic contributions in the source
term, which lead us to contact terms when we are comparing
correlation functions between $SD1$ and $ASD2$. As a byproduct we
have obtained the following dual maps: \be
\bar{f}_{\mu\nu}\leftrightarrow
\bar{e}_{\mu\nu}^*(\bar{e},\phi)=-\frac{1}{m}E_{\nu}^{\s\lambda}(\bar{e}_{\lambda\mu}+\eta_{\lambda\mu}\phi)-\frac{1}{3m}\eta_{\mu\nu}E^{\lambda\sigma}\bar{e}_{\lambda\sigma}\quad;\quad
\phi\leftrightarrow\phi\ee \no the model we have found in
(\ref{asd2}) is invariant under the gauge transformations $\delta
\bar{e}_{\mu\nu}=\p_{\mu}\xi_{\nu}^T$ and $\delta \phi =0$.
Surprisingly one can also note that the set of second order terms
in (\ref{asd2}) are (all together) invariant under the gauge
transformations: \be \delta \bar{e}_{\mu\nu}=
\p_{\mu}A_{\nu}+\p_{\nu}B_{\mu}-\frac{1}{3}\eta_{\mu\nu}\p^{\alpha}(A_{\alpha}+B_{\alpha})\quad;\quad
\delta\phi=\frac{1}{3}\p^{\alpha}(A_{\alpha}+B_{\alpha})\ee

\no besides the same second order sector is altogether free of particle content, which can be seen by means of a
hamiltonian analysis and also by studying its correspondent propagator. Then we can now use this set of terms as
mixing terms in order to construct another master action with the following structure: \be
S_{M}=S_{ASD2}(\bar{e},\phi)-S_{mixing}(\bar{e}_{\mu\nu}-\bar{f}_{\mu\nu},\phi-\chi)\ee

\no which after the shifts $\bar{f}_{\mu\nu}\to \bar{f}_{\mu\nu}-\bar{e}_{\mu\nu}$ and $\chi\to\chi-\phi$ take
us back to the $ASD2$ model thanks to the triviality of the second order sector. On the other hand we have:

\bea S_{M}&=&-\frac{1}{2}\bar{f}_{\mu\nu}\left( E^{\mu\beta}E^{\nu\alpha}+\frac{1}{3}E^{\mu\nu}E^{\alpha\beta}\right)\bar{f}_{\alpha\beta}+(\bar{f}_{\mu\nu}-
\bar{e}_{\mu\nu})\Box \theta^{\mu\nu}\chi+\chi\Box\chi+\frac{m}{2}\bar{e}_{\mu\nu}E^{\mu\alpha}\bar{e}_{\alpha}^{\s\beta}\nn\\
&+&\bar{e}_{\mu\nu}\left( E^{\mu\beta}E^{\nu\alpha}+\frac{1}{3}E^{\mu\nu}E^{\alpha\beta}\right)\bar{f}_{\alpha\beta}+3m^2\phi^2-
\phi\Box\theta^{\mu\nu}\bar{f}_{\mu\nu}-2\phi\Box\chi+\phi T\nn\\
&-&\frac{1}{m}\bar{e}_{\mu\nu}E^{\mu}_{\s\alpha}\bar{T}^{\alpha\nu}+\frac{1}{m}\phi
E_{\mu\nu}\bar{T}^{\mu\nu}.\eea

\no After functionally integrating over $\bar{e}_{\mu\nu}$ and the
scalar $\phi$ and then defining
$f_{\mu\nu}=\bar{f}_{\mu\nu}+\eta_{\mu\nu}\chi$ we arrive at an
alternative, and unusual, new self-dual model which contains
second, third and fourth order terms in derivatives:

\bea S_{ASD4}&=&-\frac{1}{2}f_{\mu\nu}\left( E^{\mu\beta}E^{\nu\alpha}+\frac{1}{3}E^{\mu\nu}E^{\alpha\beta}\right)
f_{\alpha\beta}-\frac{1}{2m}f_{\mu\nu}\Box\left(  \theta^{\mu\alpha}E^{\nu\beta}-\frac{2}{3}\theta^{\mu\nu}E^{\alpha\beta}\right)f_{\alpha\beta}\nn\\
&-&\frac{1}{12m^2}f_{\mu\nu}\Box^2
\theta^{\mu\nu}\theta^{\alpha\beta}f_{\alpha\beta}+f_{\mu\nu}^*T^{\nu\mu}\,\,\,
, \label{asd4model}\eea

\no where we have defined the dual field: \be
f_{\mu\nu}^*=\frac{1}{m^2}\left(
E_{\mu\alpha}E_{\nu\beta}+\frac{1}{3}E_{\nu\mu}E_{\alpha\beta}\right)f^{\alpha\beta}-\frac{1}{6m^3}\Box
E_{\nu\mu}
\theta_{\alpha\beta}f^{\alpha\beta}+\frac{1}{2m^2}\Box\eta_{\mu\nu}\theta_{\alpha\beta}f^{\alpha\beta}\label{asd4}\ee

\no One can check that correlation functions of $e_{\mu\nu}$ in
the first order self-dual model of \cite{aragone} coincide with
correlation functions of the dual field $f_{\mu\nu}^*$ in the
model $S_{ASD4}$ up to contact terms. The model (\ref{asd4model})
is invariant under  the gauge transformation $\delta
f_{\mu\nu}=\p_{\mu}A_{\nu} +\p_{\nu}B_{\mu}$  which leaves
$f_{\mu\nu}^*$ also invariant.

Remarkably, the model $S_{ASD4}$ can be written in the form (\ref{sdn}) with help of (\ref{asd4}). Although the
fourth order term of (\ref{asd4model}) has no particle content, we have not been able to produce any higher
(than four) selfdual model out of $S_{ASD4}$. It seems that  the highest number of derivatives in spin-2 models
in $D=2+1$ is indeed four.


\subsection{Scalar-tensor New Massive Gravities}

One way of obtaining the  New Massive Gravity of \cite{bht} is to start with the massless  linearized
Einstein-Hilbert (LEH) theory in $D=3+1$ and perform its Kaluza-Klein dimensional reduction leading to the
massive Fierz-Pauli theory in $D=2+1$ from which we obtain NMG as a dual model  via a master action technique
\cite{dj} where the mixing term between old and new (dual) fields is the full Einstein-Hilbert theory, see
\cite{sd4}. If we replace the LEH by the WTDIFF model in $D=3+1$, the KK dimensional reduction  leads to a
massive model where one of the Stueckelberg fields can not be gauged away, see \cite{bfh}. We may choose to end
up with a lower dimensional theory which corresponds to the FP model after the replacement
$h_{\mu\nu}\to\bar{h}_{\mu\nu}+\eta_{\mu\nu}\phi$. This is physically equivalent to the usual FP model since it
could have been obtained by first introducing a scalar
 Stueckelberg field $h_{\mu\nu} \to h_{\mu\nu} +
\eta_{\mu\nu}\phi $, altogether with a Weyl symmetry, followed by the harmless shift $\phi \to \phi - h/3$. This
new form of the FP theory inspires us to define a new master action with a traceless mixing term:

\bea {\cal L}_{M}&=&\frac{1}{2}(\bar{h}_{\mu\nu}+\eta_{\mu\nu}\phi)E^{\mu\alpha}E^{\nu\beta}
(\bar{h}_{\alpha\beta}+\eta_{\alpha\beta}\phi)-\frac{m^2}{2}\left\lbrack (\bar{h}_{\mu\nu}+\eta_{\mu\nu}\phi)^2-(3\phi)^2\right\rbrack\nn\\
&-&\frac{1}{2}\left(\bar{h}_{\mu\nu}- \bar{f}_{\mu\nu}\right)
E^{\mu\alpha}E^{\nu\beta}\left(\bar{h}_{\alpha\beta}- \bar{f}_{\alpha\beta}\right) \quad . \label{lm}\eea

\no Since the traceless LEH theory has no propagating degree of freedom, after the shift $\bar{f}_{\mu\nu} \to
\bar{f}_{\mu\nu} + \bar{h}_{\mu\nu} $ the fields decouple and it is clear that the particle content of
(\ref{lm}) is the same one of the massive FP model, i.e., one helicity doublet $+2$ and $-2$. On the other hand,
after integrating over $\bar{h}_{\mu\nu}$ in (\ref{lm}) we have a quadratic scalar tensor model depending upon
$(\phi , \bar{f}_{\mu\nu} )$. If we suppose that such theory comes from the singular replacement (gauge fixing
at action level) $f_{\mu\nu} \to \bar{f}_{\mu\nu}$ of a full reparametrization invariant model, its simplest
nonlinear completion  would be

\be {\cal L}_{SNMG}= 2\, \sqrt{-g}\left\lbrack - R + \frac 1{m^2}\left(R_{\mu\nu}^2 - \frac 13 R^2 \right) +
\frac 12 \phi \, r(\Box) \, R + \frac 12 \phi \, s(\Box) \, \phi \right\rbrack \label{lnl} \ee

\no where $g_{\mu\nu} = \eta_{\mu\nu} + f_{\mu\nu}$ and

\be r(\Box) = - \frac{\Box}{3\, m^2} \quad ; \quad s(\Box) = 3\,
m^2 - \Box + \frac{\Box^2}{3\, m^2} \quad . \label{rs} \ee

\no The model (\ref{lnl}) is a scalar modification of  NMG. This
becomes clearer after introducing an auxiliary scalar field which
allows us, using (\ref{rs}), to rewrite (\ref{lnl}) in the local
form:

\be {\cal L}_{SNMG}= 2\, \sqrt{-g}\left\lbrack - R + \frac 1{m^2}\left(R_{\mu\nu}^2 - \frac 38 R^2 \right) +
\chi \, (R - \Box \phi )   - \frac 32 \, m^2 \, \chi^2  - \frac 12 \phi (\Box - 3 \, m^2) \phi \right\rbrack
\quad . \label{chi} \ee

\no Following \cite{bht} we can eventually introduce an auxiliary
symmetric field  and bring (\ref{chi}) to a fully second order
form.

%



The NMG itself corresponds to (\ref{lnl}) with $(r,s)=(1,3\,
m^2)$. In what follows we perform an analysis of the analytic
structure of the linearized version of (\ref{lnl}) in search for
other viable (unitary and non tachyonic) scalar deformations of
NMG. The linearized version of (\ref{lnl}), using the more common
notation $g_{\mu\nu} = \eta_{\mu\nu} + h_{\mu\nu}$, can be
conveniently written as

 \bea {\cal{L}}&=& -\frac{h_{\mu\nu}\Box h^{\mu\nu}}{2}+\frac{h\Box h }{2}-\left(\p^{\mu}h_{\mu\nu}\right)^2+
\p^{\mu}h\p^{\alpha}h_{\alpha\mu}+h_{\mu\nu}\frac{\Box^2}{2m^2}\left(P_{ss}^{(2)}\right)^{\mu\nu\alpha\beta}h_{\alpha\beta}\nn\\
&+& A\left(\p^{\mu}\p^{\nu}h_{\mu\nu}-\Box h \right)^2+\phi s(\Box)\phi+\phi
r(\Box)\left(\p^{\mu}\p^{\nu}h_{\mu\nu}-\Box h \right)\label{40}\eea

\no where $s(\Box)$ and $r(\Box)$ are now arbitrary functions of
the d'Alembertian while $A$ is an arbitrary constant. In the case
of (\ref{rs}) we had $ A=1/12m^2$. We have used

\be \left\lbrack\frac{2\sqrt{-g}}{m^2}\left(R_{\mu\nu}^2
-\frac{3}{8}R^2\right)\right\rbrack_{hh} =
h_{\mu\nu}\frac{\Box^2}{2m^2}\left(P_{ss}^{(2)}\right)^{\mu\nu\alpha\beta}h_{\alpha\beta}
\label{pss2nmg} \ee

\no with the spin-2 and spin-0 (for later use) projection
operators given by

\be \left( P_{SS}^{(2)} \right)^{\lambda\mu}_{\s\s\alpha\beta} =
\frac 12 \left( \theta_{\s\alpha}^{\lambda}\theta^{\mu}_{\s\beta}
+ \theta_{\s\alpha}^{\mu}\theta^{\lambda}_{\s\beta} \right) -
\frac{\theta^{\lambda\mu} \theta_{\alpha\beta}}{D-1} \quad , \quad
\left( P_{SS}^{(0)} \right)^{\lambda\mu}_{\s\s\alpha\beta} =
\frac{\theta^{\lambda\mu} \theta_{\alpha\beta}}{D-1} \quad ,
\label{ps2} \ee

\no After Gaussian integrating the scalar field,
 we  rewrite the lagrangian as follows

\bea {\cal
L}_{SNMG}&=&-(\p^{\mu}h_{\mu\nu})^2+\p^{\mu}h\left\lbrack 1+2\Box
F(\Box)\right\rbrack\p^{\alpha}h_{\alpha\mu}+
(\p_{\mu}\p_{\nu}h^{\mu\nu})F(\Box)(\p_{\alpha}\p_{\beta}h^{\alpha\beta})\nn\\
&+&h\left\lbrack \frac{\Box}{2}+\Box^2F(\Box)\right\rbrack
h-\frac{h_{\mu\nu}\Box h^{\mu\nu}}{2}+h_{\mu\nu}\left(\frac{\Box^2
P_{ss}^{(2)}}{2m^2}\right)^{\mu\nu\alpha\beta}h_{\alpha\beta}\, ,
\label{lsnmg2} \eea

\no where

\be F(\Box) = A - \frac{r(\Box)^2}{4\, s(\Box)} \quad , \label{F}
\ee

\no The Lagrangian (\ref{lsnmg2}) can be further written in terms
of a four indices differential operator ${\cal L}_{SNMG} \equiv
h^{\mu\nu}G_{\mu\nu\alpha\beta}h^{\alpha\beta} $. The inverse
$G^{-1}$ does not exist due to DIFF symmetry. After adding the de
Donder gauge fixing term $ {\cal L}_{GF} = \lambda \left(
\p^{\mu}h_{\mu\nu} - \p_{\nu}h/2 \right)^2 $ and suppressing the
indices we have

\be G^{-1} = \frac{2\, m^2 \, P_{SS}^{(2)}}{\Box(\Box - m^2)}  +
\frac{2 \, P_{SS}^{(0)}}{\Box \left\lbrack 1 + 4\, \Box \, F(\Box)
\right\rbrack } + \cdots \label{g-1} \ee

\no where dots stand for terms which vanish when we saturate
$G^{-1}$ with conserved sources and build up a gauge invariant
amplitude. The NMG case is recovered at $F(\Box)=0$. The two point
amplitude in the momentum space is given by

\be {\cal A}(k) = - \frac i2 T_{\mu\nu}^*(k) (G^{-1})^{\mu\nu\alpha\beta}(k) T_{\alpha\beta}(k) \quad .
\label{ak} \ee

\no Where $G^{-1}(k) = G^{-1}(\p_{\mu} \to i \, k_{\mu} )$ and
$k^{\mu}T_{\mu\nu} = 0 $. More explicitly we have

\be {\cal A}(k) = i \left\lbrack \frac{S^{(0)}}{k^2\left\lbrack 1 - 4\, k^2 \, F(-k^2) \right\rbrack} -
\frac{m^2}{k^2(k^2+m^2)} S^{(2)} \right\rbrack \quad . \label{ak2} \ee

\no With $k^2=k_{\mu}k^{\mu}$ and

\bea
S^{(0)} &=& T_{\mu\nu}^* (P_{SS}^{(0)})^{\mu\nu\alpha\beta} T_{\alpha\beta} = \frac{\vert T \vert^2}2  \quad , \label{s0} \\
S^{(2)} &=& T_{\mu\nu}^* (P_{SS}^{(2)})^{\mu\nu\alpha\beta} T_{\alpha\beta} = T_{\mu\nu}^*T^{\mu\nu} -
\frac{\vert T \vert^2}2 \quad , \label{s2} \eea

\no where $T=\eta_{\mu\nu}T^{\mu\nu} = - T_{00} + T_{ii} $ is the trace of the source in momentum space.

 The analytic structure of $ {\cal A}(k)$
determines the particle content of the theory. Physical particles
correspond to simple poles with residues with positive imaginary
part. First we look at the massless pole $k^2=0$. Since both
$S^{(2)}$ and $S^{(0)}$ are Lorentz invariant we can choose the
convenient frame $k^{\mu} = (k,\epsilon,k)$, at the end we take
$\epsilon \to 0$. In \cite{enmg} we have shown that in such frame,
up to terms of order $\epsilon$ and higher, we may write

\be S^{(0)} = S^{(2)} =  \vert T_{11} \vert^2/2 \quad . \label{s0=s2} \ee

\no Therefore, requiring

\be \lim_{k^2\to 0} k^2 \, F(-k^2) = 0 \quad \Leftrightarrow \quad
\lim_{k^2 \to 0} \frac{k^2 \, [r(-k^2)]^2}{s(-k^2)} = 0 \quad ,
\label{lim} \ee

\no the imaginary part of the residue at $k^2 = 0$ vanishes and we
get rid of the massless pole,

\be I_0 = \Im \lim_{k^2 \to 0} k^2 \, {\cal A}(k) =  S^{(0)} -
S^{(2)} = 0 \quad . \label{i0} \ee

\no The same mechanism works in the NMG case, see \cite{oda}.

Now we look at possible massive poles $k^2= - \tmm^2 $. We choose
the rest frame $k^{\mu} = (\tmm,0,0)$. From $k^{\mu}T_{\mu\nu} = 0
$ one can show \cite{enmg} that, up to terms of order $\epsilon$
and higher,

\bea S^{(2)} &=& 2 \, \vert T_{12} \vert^2 + \frac 12 \vert T_{11}
- T_{22} \vert^2 \quad , \label{s2m} \\
 S^{(0)} &=&  \vert
T_{11}\vert^2 + \vert T_{22}\vert^2  - \frac 12 \vert T_{11} -
T_{22} \vert^2 \quad . \label{s0m}  \eea

\no We see that $S^{(2)} > 0$ while $S^{(0)}$ has no definite
sign. Thus, if we have any massive pole coming from $\left\lbrack
1 - 4\, k^2 \, F(-k^2) \right\rbrack = 0 $, with $\tmm \ne m$, its
residue will be proportional to $S^{(0)}$ and we are doomed to
have a ghost. It is impossible to have a physical massive scalar
particle with $\tmm \ne m$. The case $k^2 = \tmm^2 = m^2 $ is
subtler since the residue acquires contribution from both spin-2
and spin-0 sectors. Let us suppose that

\be 1 - 4\, k^2 \, F(-k^2)  \equiv G(k^2)(k^2 + m^2) \quad .
\label{G} \ee

\no with some continuous function $G(k^2)$. Consequently, we have
the imaginary part of the residue:

\be I_m \equiv \Im \, \lim_{k^2 \to - m^2} (k^2 + m^2){\cal A}(k)
= S^{(2)} - \frac{S^{(0)}}{m^2 G(-m^2)} \quad . \label{im} \ee

\no If we take an arbitrary real constant $a$, we see from
(\ref{s2m}) and (\ref{s0m})  that $S^{(2)} + a\, S^{(0)}
> 0$ whenever $0 \le a \le 1$, consequently we must have $G(-m^2)
\le -1/m^2$. On the other hand, from (\ref{lim}) and (\ref{G}) we
get $G(0)=1/m^2$. From those two points and the continuity of
$G(k^2)$ it is clear that $G(- b m^2)=0$ with some $0< b < 1$.
However, as we have argued before, we are not allowed to have a
massive scalar particle with $\tmm \ne m $. So (\ref{G}) can not
be true and there can not be any contribution to the residue at
$k^2=-m^2$ coming from the denominator of $S^{(0)}$ in ${\cal
A}(k)$. Thus, we are left with $I_m = S^{(2)} > 0$ and we are left
with only one massive spin-2 particle in the spectrum just like
the NMG case.

The previous arguments amount to require that the numerator of the
function

\be H(\Box) \equiv 1 + 4 \, \Box \, F(\Box) = \frac{\left(1 + 4 \,
A \, \Box \right) s(\Box) - \Box \, [r(\Box)]^2}{s(\Box)} \quad .
\label{H} \ee

\no be independent of $\Box$. Thus, the polynomials $r(\Box)$ and
$s(\Box)$ must be such that

\be [r(\Box)]^2 = 4\, A \, s(\Box) + \frac{[ s(\Box) - s_0
]}{\Box} \quad . \label{r2} \ee

\no Where $A$ is an arbitrary constant and $s_0 = s(\Box =0)$

After integration over the scalar field in (\ref{lnl}) using (\ref{r2}), we have the following class of spin-0
nonlocal deformations of NMG:

\be {\cal L}_{NL-NMG} = - \sqrt{-g} \, R + \frac 1{m^2}  \sqrt{-g}
\left( R_{\mu\nu}^2 - \frac 38 R^2 \right) -  \sqrt{-g} R
\frac{\left\lbrack s(\Box)-s_0\right\rbrack}{8\, \Box \, s(\Box)}
R \quad . \label{nlmmg} \ee

\no The case $s(\Box)=s_0$ corresponds to the NMG \cite{bht}. The reader can check that $r(\Box), s(\Box)$ and
$A$ given in (\ref{rs}) and in the text after (\ref{40}) respectively, satisfy (\ref{r2}).

Another special case is $s_0 =0$ where the function $H(\Box)$
vanishes. Such momentum independent singularity in $G^{-1}$
indicates the presence of a spin-0 local symmetry, in fact we have
a Weyl symmetry. The corresponding model has been found before in
our previous work \cite{enmg}. It corresponds to make the
Stueckelberg substitution $h_{\mu\nu} \to h_{\mu\nu} +
\eta_{\mu\nu} \phi $ in the LNMG and then build up its simplest
nonlinear completion. Since this is not equivalent to first take
the nonlinear NMG and then make $g_{\mu\nu} \to e^{\phi}
g_{\mu\nu}$, we expect that the linearized unitarity of the $s_0
=0$ case breaks down at nonlinear level, since $\phi$ stops being
a pure gauge degree of freedom at nonlinear level, so the Weyl
symmetry only exists in the linear theory.

Regarding the other solutions to (\ref{r2}), since they are not
associated with any local symmetry it is not clear whether the
unitarity of the linearized model is broken in the nonlinear
theory (\ref{lnl}).

\section{Conclusion}

Here we have examined different issues regarding the Weyl and
transverse diffeomorphism (WTDIFF) symmetry in $D=2+1$ massive
spin-2 theories as well as their nonlinear analogues (unimodular
theories).

Although WTDIFF theories correspond to gauge fixed versions of DIFF theories, the issue of gauge fixing at
action level is nontrivial, see \cite{mst}. In particular, the triviality of Einstein-Hilbert gravity in $D=2+1$
is lost in its unimodular $(g=-1)$ version as we have briefly commented in the beginning of section II using the
linearized theory. Instead of flat space we have now  a maximally symmetric space in general which may include
BTZ black holes \cite{btz} in the nonlinear case, depending on the sign of an integration constant which plays
the role of a cosmological constant.

We have explicitly checked that WTDIFF versions of massive spin-2
theories (one and two helicities) are fully consistent. In the
special cases of the third and fourth order (in derivatives)
selfdual (one helicity) theories, they correspond to linearized
versions of unimodular topologically massive gravity and
unimodular higher derivative topologically massive gravity.
Likewise, in the case of a parity doublet we have a linearized
version of a unimodular New Massive Gravity.

At the end of section II we have examined the issue of
renormalizability and Weyl symmetry. We argue that although both
highest and lowest derivative terms in the free (quadratic) sector
of unimodular TMG and unimodular NMG are Weyl invariant, we still
have a mismatch of local symmetries which is dangerous for
renormalizability as pointed out in \cite{deserprl}. The
Einstein-Hilbert theory is only invariant under WTDIFF (linearized
theory) while the highest derivative term (gravitational
Chern-Simons term or the the K-term) is invariant under full
WDIFF. Thus, the pure longitudinal degree of freedom $h_{\mu\nu}
\sim \p_{\mu}\p_{\nu} \phi $ only appears in the Einstein-Hilbert
term. Consecutively, it propagates like $1/p^2$ at large momentum
and no improvement is achieved for renormalizability in unimodular
theories. The mismatch between the symmetries of the highest
derivative term and the lower one seems to be unavoidable. In
\cite{annals} we have pointed out that there exists a massive
spin-2 model in $D=2+1$ described by a nonsymmetric tensor
$e_{\mu\nu}$, see \cite{jm}, where both the second and fourth
order terms are Weyl invariant, however only the fourth order one
is invariant under antisymmetric shifts. The mismatch also occurs
in the higher dimensional analogue of the linearized NMG, see
\cite{bfmrt} and \cite{ds}. This is the higher derivative analogue
of the usual breakdown of gauge symmetries by mass terms as in the
Proca (s=1) and massive Fierz-Pauli (s=2) theories. The only hope
is to find a theory where the lowest derivative term has  already
more than two derivatives.

It is known that massive theories in $D$ dimensions can be obtained from $D+1$ massless  theories via
Kaluza-Klein dimensional reduction. From the massless spin-2 linearized Einstein-Hilbert theory in $D=3+1$ one
can obtain the massive spin-2 Fierz-Pauli theory in $D=2+1$. From the later theory one can derive, via the
master action approach of \cite{dj}, the fourth order linearized New Massive Gravity theory \cite{bht}. A key
point in this approach is the absence of propagating degrees of freedom of the Einstein-Hilbert theory in
$D=2+1$ which works like a mixing term between old and new (dual) fields in the master action approach. If
however, we replace the linearized EH theory in $D=3+1$ as starting point by the the WTDIFF massless spin-2
theory, its dimensional reduction\footnote{In \cite{jm} the NMG theory has been directly obtained via
Kaluza-Klein dimensional reduction from the LEH theory where the scalar Stueckelberg has been eliminated in a
unusual way. We are currently investigating a similar procedure applied to the $D=3+1$ WTDIFF theory.}
\cite{bfh} leads to the massive FP theory with the replacement $h_{\mu\nu} \to \bar{h}_{\mu\nu} + \eta_{\mu\nu}
\phi$. In section III, starting from the latter theory we have defined a noncanonical (traceless) master action
where the mixing term is the EH action for the traceless field $\bar{h}_{\mu\nu}$. This leads us to a scalar
tensor modification of the NMG theory. We have shown it belongs to a more general class of consistent (unitary
at quadratic level) scalar tensor modifications of NMG. The consistency of their nonlinear completion
(\ref{lnl}) demands further investigations.

From the point of view of dimensional reduction the appearance of
an extra scalar field in the massive model is a consequence of the
fact that the gauge parameter of the massless higher dimensional
theory satisfies the scalar constrain $\p^{M} \xi^T_{M}=0$ with
$M=0,\cdots , D$. This makes the lower dimensional gauge
parameters not independent, consequently we are not able to
eliminate all the Stueckelberg fields and we may choose to remain
with one scalar Stueckelberg field, see \cite{bfh}. This is
similar to the massive spin-3 theory which requires an extra
scalar field besides a totally symmetric rank-3 tensor
$\phi_{\alpha\beta\gamma}$ due to the constrained symmetry of the
higher dimensional massless theory
$\delta\phi_{\alpha\beta\gamma}=\p_{(\alpha}\bar{\xi}_{\beta\gamma
)}$ where $\eta^{\mu\nu}\xi_{\mu\nu}=0$.

Still in section IV we have also applied the noncanonical master action approach on the first order selfdual
model of \cite{aragone} and derived a new second order model (NSD2) which, on its turn, has given rise to a new
fourth order model (NSD4). The unusual NSD4 model contains second, third and fourth order terms. Remarkably, the
SDn models and also NSD4 can all be written in the compact form (\ref{sdn}) which also works in the spin-1 case
(see footnote (3)). We believe that such compact formulas may exist also for higher spins which might help us in
filling some gaps in the chain of spin-3 and of even higher spin selfdual models.

\section{Acknowledgements} The works of D.D. and E.L.M. are
partially supported by CNPq under grants (307278/2013- 1) and (449806/2014-6) respectively. The authors would
like to thank FAPESP grants 2011/11973-4 and 2016/01343-7, for funding the visit to ICTP-SAIFR where part of
this work was done.

\end{document}